\documentclass[fleqn,twoside]{article}
\usepackage{espcrc2}
\usepackage{epsfig}

\usepackage{graphicx}
\usepackage[figuresright]{rotating}

\newcommand{\AmS}{{\protect\the\textfont2
  A\kern-.1667em\lower.5ex\hbox{M}\kern-.125emS}}

\title{Tile and crystal calorimeters for the KLOE$^2$ experiment}

\author{F.Happacher\address{Laboratori Nazionali di Frascati dell'INFN},
        M.Martini$^{a,}$\address{Dipartimento 
di Energetica Univ. Roma La Sapienza}\thanks{Corresponding author. 
Mail: matteo.martini@lnf.infn.it},
        S.Miscetti$^{a}$,
        I.Sarra$^{a}$}
       
\begin{document}

\begin{abstract}

The upgrade of the DA$\Phi$NE machine layout requires a modification
of the size and position of the inner focusing quadrupoles of 
KLOE$^2$ thus asking for  the realization of two new calorimeters 
covering the quadrupoles area.
To improve the reconstruction of $K_L\to 2\pi^0$ events 
with photons hitting the quadrupoles, a tile calorimeter, QCALT, 
with high efficiency to low energy photons (20-300 MeV),
time  resolution of less than 1 ns  and space resolution of few cm,
is needed. We propose a tile calorimeter 
with a high granularity readout corresponding to about
2500 silicon photomultipliers (SiPM) of $1\times 1$ mm$^2$ area. 
Moreover, the low polar angle regions need the realization of
a dense crystal calorimeter  with very high time resolution 
performances to extend the acceptance for multiphotons events. 
Best candidates for this calorimeter are LYSO crystals with
APD readout or PbWO$_4$ crystals with large area SIPM readout.


\vspace{1pc}
\end{abstract}

\maketitle

\section{The KLOE$^2$ proposal}

In the last decade a wide experimental program has
been carried out at Da$\Phi$ne\cite{dafne}, the $e^+e^-$
collider of the Frascati National Laboratories,
running at a center of mass energy of 1020 MeV, the 
$\Phi$ resonance mass. During KLOE run, Da$\Phi$ne
delivered a peak luminosity of 1.5$\times$10$^{32}$
cm$^{-2}$s$^{-1}$ 
which granted about 1 fb$^{-1}$ per year in the last 
data taking period. 

A new machine scheme has been recently proposed 
by the Frascati accelerator group aiming at increasing
the luminosity of the machine up to a factor 5.
This scheme has been succesfully tested at Da$\Phi$ne,
and these encouraging results push for 
a new data taking compaign
for the KLOE experiment to complete its physics program 
and to perform a new interesting set of measurements. 

The new experiment, named KLOE$^2$,
expects to collect 5 fb$^{-1}$/year.
We are now working to 
improve the performances
of our detector\cite{kloe_all} adding: an inner tracker, a tagger
system to study $\gamma\gamma$ physics, a new small angle calorimeter
and a new quadrupole
calorimeter.
In this paper we explain the project and the 
R\&D for these last two items.

\section{Quadrupole tile calorimeter, QCALT}

\begin{figure}[htb]
\psfig{file=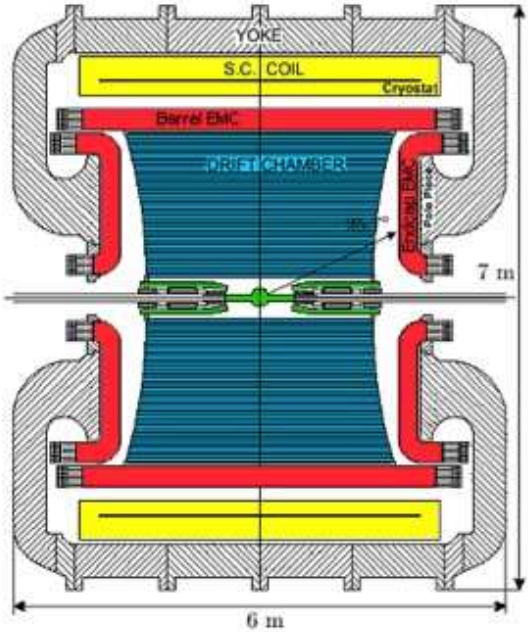,width=8cm}
\vspace{-1.5cm}
\caption{Section of the KLOE detector.}
\label{figkloe}
\end{figure}

In Fig.\ref{figkloe}, we show a section of the KLOE detector in which 
is visible the old position of the focalizing quadrupoles and 
the surrounding
calorimeters QCAL \cite{oldqcal}
which have a polar angle 
coverage of 0.94$\,<|\cos\theta|<\break$0.99.
Each calorimeter consists of  16 azimuthal sectors composed
by alternating layers of 2 mm lead and 1 mm BC408 
scintillator tiles, for a total thickness of $\sim$5X$_0$.
The readout is done by wavelength shifter fibers
 (Kuraray Y11-200) and photomultipliers. The fiber arrangement allows
the measurement of the longitudinal coordinate by time differences.
These calorimeters are characterized by a low light response 
($\sim$3 pe/mip/tile at zero 
distance from the photomultiplier) due 
to the  coupling in air, to the fiber lenght 
($\sim$2 m for each tile) and 
to the quantum efficiency of the used photomultipliers (standard bialkali 
with $\sim$20\% QE).

The project of the new QCAL consists in a dodecagonal structure,
1 m long, covering the region of
the quadrupoles. The structure
consists in a sampling of 5 layers of 
5 mm thick scintillator plates alternated
with 3.5 mm thick tungsten plates,
for a total depth of 4.75 cm (5.5 X$_0$). 
The active part of each plane is divided into 
twenty tiles of 5$\times$5 cm$^2$ area 
with 1 mm diameter WLS fibers embedded 
in circular grooves. Each fiber is then
optically connected to a silicon 
photomultiplier of 1 mm$^2$ area, SiPM, 
for a total of 2400 channels.

We have done some R\&D studies on SiPM, fibers and tiles 
to choose the combination which optimizes 
the response of our system.

\subsection{Test on MPPC}

We have compared the characteristics of two different SiPM 
produced by Hamamatsu (multi pixel photon counter, MPPC):
100 (S10362-11-100U) and 400 pixels 
(S10362-11-050U), both with 1$\times 1$ mm$^2$ active area. 
To manage the signals, the electronic 
service of the Frascati Laboratory (SELF) has 
developed a custom electronics composed by
a 1$\times$2 cm$^2$ card, containing the pre-amplifier, and 
a multifunction NIM board. 
For these tests, we have set the
pre-amplifier gain to 50. The NIM board 
supplies the voltage to the photodetector (Vbias) with 
a precision of 2 mV and a stability at the level of 
0.03 permill. A low threshold 
discriminator and a fanout are also present in the board.

To determine the gain, we have prepared a setup 
based on a blue light LED and a polaroid filter to change 
light intensity.
We have measured the gain and the dark rate
variation as a function of the applied Vbias
and the temperature of the photodetector.

The readout electronics was based on CAMAC, 
with a charge sensitivity of 0.25 pC/count
and a time of 125 ps/count. 

Our tests confirm the performances declared by Hamamatsu and
show a significative variation of the detector gain 
as a function of the temperature. 
The 400 pixels shows a temperature dependence of the gain which is 
a factor four smaller than the 100 pixels (3\% versus
12\%), with a gain reduction of 
a factor five.

\subsection{Tests on fibers}

We have studied the light response of three different, 
1 mm$^2$, fibers optically connected to MPPC:
\begin{itemize}
\item Kuraray SCSF81 (blue scintillating)
\item Saint Gobain BCF92 single cladding 
(WLS from blue to green)
\item Saint Gobain BCF92 multi cladding 
(WLS from blue to green)
\end{itemize}

The test is done firing the fiber with a Sr$^{90}$ source.
The trigger is provided by a NE110 scintillator finger 
(1$\times$5$\times$0.5 
cm$^3$) connected to a bialkali photomultiplier positioned below
the fiber. 

\begin{figure}[htb]
\psfig{file=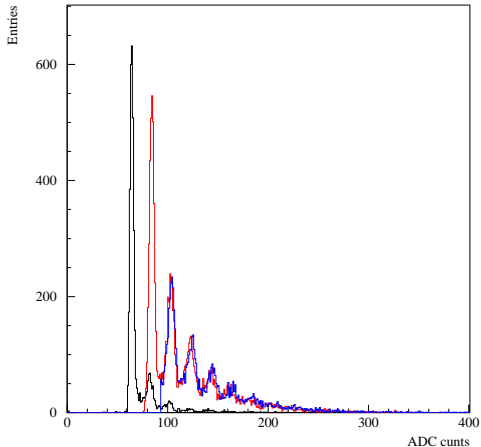,width=7cm}
\vspace{-1cm}
\caption{Charge response to Sr$^{90}$ for a BCF92
multicladding fiber.The different
spectra are obtained setting threshold at 0.5, 1.5 and 2.5 
photoelectrons.}
\label{figsgmc}
\end{figure}

As expected, a large light yield is shown for SCSF81 while the WLS fibers
have a reduced response. However, the BCF92 multi cladding has a 
reasonable light yield as shown in Fig.\ref{figsgmc}. 
For this fiber we have: maximum
light yield, fast response (5 ns/pe) and high 
attenuation length (3.5 m).

\subsection{Tests on tiles} 

\begin{figure}[htb]
\vspace{9pt}
\psfig{file=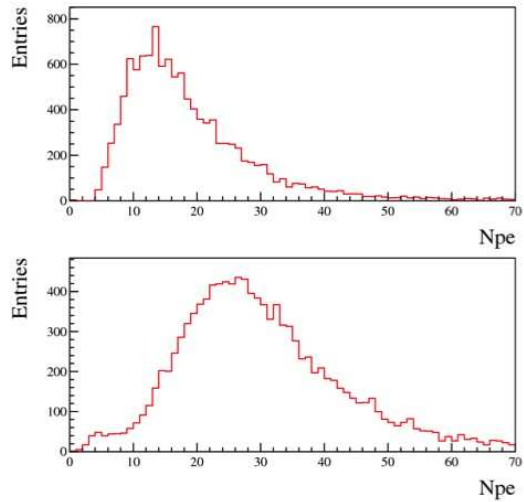,width=7cm}
\vspace{-1cm}
\caption{ADC distribution, in number of photoelectrons,
for two different type of tiles.}
\label{figtile}
\end{figure}

Light response and time resolution of tiles have been 
measured using cosmic rays. The system was prepared connecting
the fiber to the MPPC and using two NE110 fingers to trigger 
the signal. We have prepared two different tiles:
\begin{description} 
\item[A]  3 mm thick tile with 400 pixels MPPC,
\item[B]  5 mm thick tile with 100 pixels MPPC.
\end{description}
Using ADC distribution we find: 14 pe/mip for tile ``A'' and
26 pe/mip for tile ``B'' 
(See Fig.\ref{figtile}). These results
are comparable taking into account the thickness ratio between tiles
and the photon detection efficiency of the two
detectors (40\% for 400 pixels and 45\% for 100 pixels). 

Correcting for the time dependence
on pulse height, we find a preliminary
time resolution of 1000 ps (750 ps) for tile ``A'' (``B'').
 
\subsection{Next plans}

We are now assembling two small 
dimension multi-tiles 
prototypes of the QCAL, to study
signal transportation and to 
measure the effective radiation length. 
In 2009,
we plan also to construct a ``module 0''
consisting of a complete slice of the dodecagon 
(1/12 of one calorimeter) with final material and electronics. 

\section{Crystal calorimeter with timing, CCALT}

In the new design of Da$\Phi$ne 
interaction region, the
position of the quadrupoles increases the acceptance
of the central calorimeter from 21$^{\circ}$ to
18$^{\circ}$. Below this limit we can safely insert
few crystals to improve the acceptance for photons
coming from $\eta$ and $K_S$ decays. This detector could
work as veto detector for photons 
down to 8$^{\circ}$. The particular region is visible in
Fig.\ref{figkloe} and is delimited between the 
focalizing quadrupoles and the spherical interaction region
of the KLOE detector.

The proposed solution is to insert an
homogeneus calorimeter based on LYSO 
($Lu_{18}Y_{.2}SiO_5:Ce$) crystals.
The most important characteristics
of these crystals are a very high light yield,
a time emission, $\tau$, of 40 ns,
high density and X$_0$ without beeing
hygroscopic (See Tab.\ref{tablfs}).
These crystals well match the request
of high efficiency to low energy
photons and excellent time resolution for
the CCALT, which will help in 
fighting the high level of machine
background events present in the low energy
region.

The preliminary project consists in a dodecagonal
barrel for each side of the interaction region,
composed by LYSO crystals (2$\times$2$\times$13 cm$^3$)
readout using
avalanche photodiodes (APD). 

\subsection{Preliminary tests on LYSO}

We have tested one LYSO crystals from Saint Gobain
(2$\times$2$\times$15 cm$^3$) 
using both
cosmic rays and electrons from Frascati beam
test facility (BTF).
For cosmic rays test, the crystal was
readout using standard bialkali photomultiplier, 
while in the test beam
we used a 5$\times$5 mm$^2$ avalanche 
photodiode provided 
by Hamamatsu (S8664-55).

In both cases, the crystal was simply wrapped with mylar and
the photodevice coupled with optical grease and
simple mechanical arrangement.  

After correcting the time dependence on 
pulse height,
we obtain, from cosmic rays, a preliminary time resolution 
$\sigma_T=360$ ps,
which corresponds to 12000 pe/mip
($N_{pe}=(\tau/\sigma_T)^2$).
Assuming a MIP to deposit 10 MeV/cm 
this correponds to 600 pe/MeV. 

At the BTF, we have used
electrons from 50 to 400 MeV to
measure the time resolution 
of the crystals which is well 
parametrized by:
$$
\sigma_T=\frac{82\,ps}{\sqrt{E(GeV)}}\oplus 293\,ps
$$

The constant term is probably related to the different 
arrival times of the showering photons along
the crystal axis.
From the statistical term we derive a light yield 
of 240 pe/MeV which is in rough agreement with 
the results from cosmic rays taking into account
quantum and collection efficiency.

\subsection{Next plans}

A dedicated test beam 
with electrons will be held in Frascati beam test
facility in the first
months of 2009. 
The aim of the test is 
to characterize energy and time 
resolution of a 3$\times$3 matrix of
high quality crystals surrounded by
an outer leakage detector
done with PbWO.
This test will also allow to compare the response
of different kind of LYSO crystals (Saint Gobain 
and Scionix) and a similar crystal (LFS by
Zecotek) which could be an alternative candidate
(See Tab.\ref{tablfs}).

\begin{table}
\begin{tabular}{|c|c|c|}
\hline
 & LYSO & LFS \\
\hline
Density & 7.1 & 7.2-7.3 \\
Attenuation lenght (cm) & 1.2 & 1.12 \\
Decay constant (ns) & 41 & 35-36 \\
Max emission (nm) & 420 & 435-438 \\
Light yield (relative NaI) & 75 & 80-85 \\
Energy resolution & 8 & 9-12 \\
Hygroscopic & no & no \\
Refractive index & 1.81 & 1.78 \\
\hline
\end{tabular}
\caption{Comparison between LYSO and LFS characteristics}
\label{tablfs}
\end{table}

\section{Conclusions}

The new scheme proposed for the Da$\Phi$ne machine allows
a factor 5 increase in the delivered luminosity. 
Some R\&D are in progress to add new 
components to the KLOE apparatus. We have presented the
proposal for two low angle calorimeters. QCALT is a tile 
calorimeter surronding the focalizing
quadrupoles to increase the coverage of the 
electromagnetic calorimeter. CCALT is a LYSO calorimeter
aimed to be a good veto for low angle photons. 
We have presented the preliminary measurement on tiles and 
crystals and the characterization of the 
photodevices used with tiles.

\section{Acknowledgement}

We want to thank G.Corradi, D.Tagnani and C.Paglia to 
have developed the readout electronics of the SiPM; M.Cordelli
for his help during the preparation of test setup; 
M.Arpaia, G.Bisogni, A.Cassar\`a, A.Di Virgilio, U.Martini
and A.Olivieri for their help in the mechanical preparation
of the setup and in the 
preparation of the tiles.

\end{document}